# How to Assess the Carbon Footprint of a Large-Scale Physics Project

Clarisse Aujoux (a), Odile Blanchard (b) and Kumiko Kotera (a,c)
(a) Sorbonne Université, UPMC Univ. Paris 6 et CNRS, UMR 7095, Institut d'Astrophysique de Paris, 98 bis bd Arago, 75014 Paris, France

(b) Université Grenoble Alpes, GAEL Laboratory (UMR CNRS 5313, UMR INRAe 1215), CS 40700 - 38058 Grenoble CEDEX 9, France

(c) Vrije Universiteit Brussel, Physics Department, Pleinlaan 2, 1050 Brussels, Belgium

corresponding author: Kumiko Kotera <kotera@iap.fr>

**Standfirst:** Large-scale projects have become increasingly important in physics. They are also a source of greenhouse gas emissions. Clarisse Aujoux, Odile Blanchard and Kumiko Kotera describe how to use transparent, open data to estimate these emissions — the first step in taking effective action to reduce them.

Large-scale experiments are building blocks of the physics community: they involve a large fraction of the scientific staff working in multiple countries, and absorb a significant volume of the science budget. They are also a collection of carbon-emitting sources and practices. As such, it is essential to assess their environmental impact.

We describe here a methodology to estimate the main greenhouse gas (GHG) emissions of a large-scale astrophysics collaboration project, using transparent open data. The goal is neither to consider all possible emission sources of a project, nor to calculate accurate values. It is rather to identify the biggest emission sources of the project, obtain orders of magnitude for them and analyse their relative weights. We discuss methods to quantify the GHG-generating activities and their related emission factors for the three typical biggest emission sources that can be controlled by the collaboration: travel, digital and hardware (summarized in Table 1).

**Global methodology**

The first step is to set the study scope. This step aims at listing the experimental stages of the project and the persons involved in the collaboration over the years and decades, and at identifying in advance the most emitting sources of the project. Identifying who is strongly, moderately or lightly involved in the collaboration is needed to allocate an appropriate share of their annual GHG emissions to the project.

GHG emissions of a specific activity are estimated by multiplying the quantity of activity by its emission factor, i.e. the emissions generated per unit of activity. The carbon footprint of the project results from the aggregation of the GHG emissions across all the project activities.

**Quantifying emissions-generating activities**

Travel emissions stem from trips to attend meetings, conferences, on-site visits and the like. Emissions are assessed through the distances travelled and the modes of transportation. Data can be collected by surveying the collaborators about their project-related trips. From the trip description, the distances can be assessed using online calculators. Extrapolations taking into account the involvement of the collaborators can be made if the survey is incomplete.

Emissions from digital technologies stem from the devices used, communication flows, numerical simulations, data transfer and storage. A typical set of devices is defined for each collaborator, then the number of devices is scaled to the number of collaborators and their degree of involvement. For communication flows, the data rely on an estimation of how many emails are sent to and by a member of the collaboration daily and how many days are worked over the year, weighted by the degree of involvement in the collaboration. Emissions associated to video conferencing can be neglected [1]. For numerical simulations, the data required is the number of central processing unit (CPU) hours used by the collaboration; computing centres can provide this information. The quantity of data transferred and stored is assessed via the volume of raw data collected by the experiment and transmitted each day. For data storage, the data volume and the number of centres where it is backed up are the main drivers of emissions, through the electricity consumption of the cloud servers. The latter can vary greatly according to the assumptions made and the year to which the data apply [2].

The GHG emissions generated by the hardware stem from the pieces that make up the whole experiment. For example, metal parts and batteries are quantified by weight, and solar panels are accounted for in terms of installed area and the associated electricity generated. If the experiment is located far from the production site of the hardware, the distance and mode of freight transport are also assessed.

**Emission factors**

Emission factors are crucial to estimate the carbon footprint of a project. Choosing and sharing reliable open data sources for emission factors enables future comparisons of carbon footprints of other projects. If possible, using the same database for all emission factors ensures results are consistent. For an international project, one should also identify where the GHG emissions are generated, because some emission factors depend on the energy mix and manufacturing practices, which vary around the world.

The [ADEME Carbon database](#) is a good choice of database as it is free, reliable, comprises numerous emission factors including some location-specific ones, and shows the degree of uncertainty of the emission factors. However, it does not provide all the emission factors needed. In this case, one has to search for data from other sources.

Some issues related to the choice of the emission factors must also be addressed. For travel, emission factors are expressed in $CO_2$ equivalent per passenger-kilometre ($CO_2$e/pkm) for each transportation mode. Those pertaining to flights may vary by a factor of 5 between databases, owing to different methodologies and some databases not including all types of aviation emissions. Combining the figures from several databases into an average value avoids excessively high or low estimates [3].

On the digital side, emission factors of devices are available in the [EcoDiag](EcoDiag) database, including emissions due to manufacturing and usage. Emission factors have been published for a 1 MB email, including an attached file [4], and for a CPU-hour that includes all the emissions associated with a data centre: emissions from server manufacture, electricity consumption of the servers and the professional emissions of the data centre staff [5]. A fundamental issue for computing-related emission factors is which value to consider for electricity consumption, which depends on location. For example, the electricity emission factor is 0.42 kg$CO_2$e/kWh in Europe, but 0.766 kg$CO_2$e/kWh in China (data from the ADEME Carbon database). The location of cloud servers used for storage is usually unknown, but the emission factor of electricity linked to them can be estimated by calculating a weighted value based on the worldwide distribution of data centre locations.

For hardware equipment, emission factors arise from the manufacturing of the components, such as the metals of the structure, the batteries and the solar panels. These emission factors can be retrieved from databases such as the ADEME Carbon database or specific studies [6].

**Projections**

Large-scale experiments often evolve over decades. Projections can be made using linear scaling rules. At first order, GHG-generating activities scale with either the size of the collaboration (as travel, devices, communication and simulations do), or the size of the experiment (as data volume and hardware equipment number and/or size do). The foreseen growth rate of these two parameters drives the projections. Projections bear large uncertainties due to unforeseen experimental, technological and also human societal changes. Consequently, they should be viewed as order-of-magnitude estimates.

**Putting the method into practice**

This methodology has been applied to the GRAND project, a multi-stage experiment in its prototyping phase, which aims at detecting ultra-high-energy neutrinos with an array of 200,000 radio antennas over 200,000 km$^2$ [7]. Travel, digital and hardware emissions have different impacts at the different stages and scale of the experiment. The importance of data transfer and storage as GHG-generating activity was also highlighted.

Several actions will be taken, including: strategies to store data more efficiently, hence reducing its volume, in countries with low energy electricity emission factors; systematically checking hardware part production conditions, and improving the electronics design to reduce energy consumption (implying smaller batteries and solar panels, and possibly less metal); incentives to weigh the cost/benefit of numerical simulations; initiatives to hold half the collaboration meetings online, and to favour local collaborators for on-site missions.

**Acknowledgements**

We are grateful to the GRAND Collaboration. This work was supported by the Institut d'Astrophysique de Paris, and by the APACHE grant (ANR-16-CE31-0001) of the French Agence Nationale de la Recherche.

**Competing interests**
The authors declare no competing interests.

**Table 1:** Main sources of GHG emissions in a large-scale physics experiment.

| Sources of emissions | Quantity of GHG-generating activity | Emission factor |
|---|---|---|
| Professional travel | Total distances traveled per year per transportation mode | $CO_2e$ per passenger-kilometer |
| Digital | | |
| Devices (such as computer, screen) | Manufacturing and usage of each type of device | $CO_2e$ per device |
| Communication | Emails sent | $CO_2e$ per MB |
| Simulations | CPU hours | $CO_2e$ per CPU-hour |
| Data transmission and storage | Electricity consumption of servers | $CO_2e$ per kWh |
| Hardware | | |
| Metal devices | Metal weight | $CO_2e$ per kg of metal |
| System batteries | Battery weight | $CO_2e$ per kg of battery |
| Solar panels | Installed area | $CO_2e$ per m$^2$ |
| Transportation | Distances travelled and weight hauled per transportation mode | $CO_2e$ per ton-km |